\documentclass[journal]{IEEEtran}
\ifCLASSINFOpdf

\else

\fi
\usepackage{multirow}

\usepackage{graphicx}

\usepackage{color}

\begin{document}
\title{Security by Design Issues in Autonomous Vehicles}

\author{Martin~Higgins$^*$, 
        Devki Nandan Jha$^{\S,*}$, 
        David Blundell$^\psi$ 
        and~David~Wallom$^*$ 
\thanks{$^*$ Author is with Oxford e-Research centre, University of Oxford, Holywell House, 64 Osney Mead, Oxford, OX2 0ES, United Kingdom. 
     $^{\S}$ Author is with Newcastle University, 1 Science Square, Newcastle Upon Tyne, NE4 5TG, United Kingdom.
    $^{\psi}$ Author is with CyberHive Ltd., Newmarket House, Newbury, RG14 5DP, United Kingdom.}
}
\markboth{IT PROFESSIONAL MAGAZINE}%
{Shell \MakeLowercase{\textit{et al.}}: Bare Demo of IEEEtran.cls for IEEE Journals}
\maketitle
\begin{abstract}
 As autonomous vehicle (AV) technology advances towards maturity, it becomes imperative to examine the security vulnerabilities within these cyber-physical systems. While conventional cyber-security concerns are often at the forefront of discussions, it is essential to get deeper into the various layers of vulnerability that are often overlooked within mainstream frameworks. 
Our goal is to spotlight imminent challenges faced by AV operators and explore emerging technologies for comprehensive solutions. This research outlines the diverse security layers, spanning physical, cyber, coding, and communication aspects, in the context of AVs. Furthermore, we provide insights into potential solutions for each potential attack vector, ensuring that autonomous vehicles remain secure and resilient in an evolving threat landscape.
\end{abstract}

\begin{IEEEkeywords}
Autonomous vehicles, security by design, system security, cyber-physical.
\end{IEEEkeywords}
\IEEEpeerreviewmaketitle
\section{Introduction}
\IEEEPARstart{A}{s} of $2020$, the autonomous car market was valued at USD $1.45$ billion and projected to grow substantially over the next decade with some expecting the market to grow to $800$ billion by $2035$ \cite{Lanctot2017AcceleratingEconomy}. Despite the huge potential of autonomous vehicles (AV) there is trepidation about their roll-out into commercial environments. The inherent risks associated with AVs, including potential loss of life and financial repercussions, continue to hinder their widespread commercial adoption \cite{Kim2021CybersecurityDefenseb}. Operators are worried not just about operational failures but also deliberate attacks on AVs which may pose substantial security concerns.
Given the obvious danger of an out-of-control AV, cyber-security of AVs is a growing concern  \cite{Kukkala2022RoadmapVehicles}. 

\subsection{Motivation, Objectives, and Contribution} 

In this work, we attempt to outline a full design-stack security view of AVs. We proceed layer by layer across the sensing, processing, coding and encryption layers of an AV and outline some fundamental security questions. We make suggestions on how these vulnerabilities could be fixed and aim to provide a high-level overview of what cyber-physical security means for contemporary AVs. We also outline some potential solutions in the layer for each vulnerability type. By reviewing this document the reader should have a broad understanding of the type, location and potential solution of a wide range of cyber-physical vulnerabilities to modern AVs.

\section{Cyber-Physical System Stack}
System security has often been considered in the context of the individual components of the system. However, authors are increasingly taking a holistic approach to security that includes both the cyber, physical and communication layers. A cyber-physical system is one which exhibits a hierarchy of interaction between analogue, continuous signals and the quantized 0 \& 1s interpretable to a computer. Often the physical layer refers to an analogue sensing system often upstream of the cyber network. However, it can also refer to the downstream physical elements of a device which are dependent on continuous measurement phenomena such as the actuators in a robotic arm.

\begin{figure*}
    \centering
    \includegraphics[width=1.0\textwidth]{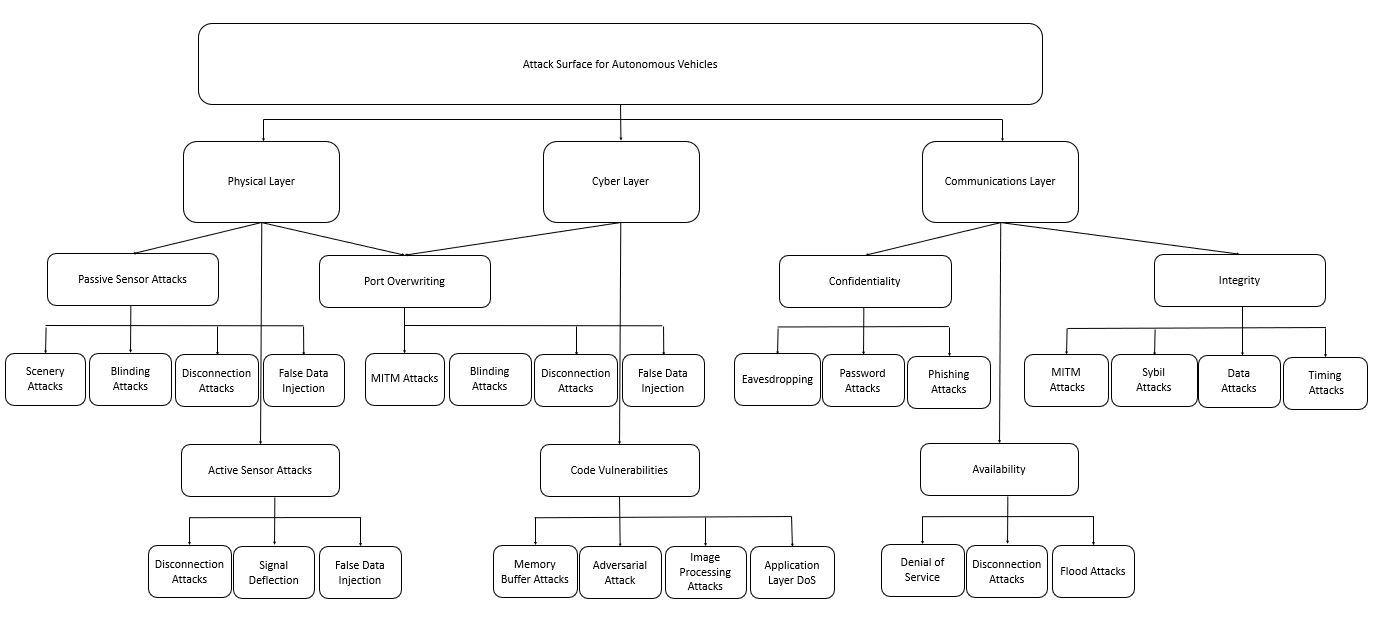}
    \caption{Taxonomy of potential attacks for the Autonomous vehicles}
    \label{fig:taxonomy}
\end{figure*}

Conventional cyber-security has understandably received much more direct focus than physical layer-style security. However, we still believe there are some under-served areas within this part of the network stack that could use some due focus. There is also the encrypted communications layer which dictates interactions with the AV central authority. In this work we examine cyber, physical and communication network vulnerabilities. In figure \ref{fig:taxonomy} we outline the attack types outlined and discussed within this work.

\section{Physical Layer Attacks}
Often we consider cyber-security in terms of the cyber layer only. However, physical later attacks can often bypass protections such as authentication, encryption and other security features outright by attacking the infrastructure upstream of these defensive processes. Here we outline some examples of physical layer processing attacks and some proposals for methods to solve these attacks.  

\subsection{Passive Sensor Attacks} 
Increasingly, AVs are reliant on visual data from camera sensors for travel. In fact, one of the leaders in the field, Tesla boasts about their reliance on cameras over other methods as they provide a generalized solution to self-driving vehicle navigation. Camera spoofing attacks have already shown to be possible for AVs \cite{PetitRemoteLiDAR}. At the most simplistic level, \emph{Blinding attacks} can be performed using a diode point laser by focusing the laser on the camera. Depending on how the AV responds to this kind of scenario the vehicle might stop leaving it vulnerable to theft or further manipulation by the attacker. AVs have also been shown to be susceptible to \emph{False classification}. In the past, self-driving cars have been fooled by billboards, fast food signs and other environmental features (including the moon). Attackers can replicate these events by creating fictional images in the camera's view. These kind of attacks are sometimes known as \emph{Scenery attacks} as the attacker changes the background scenery to force a response from the car. Also, images which appear normal to the passenger can be subtly altered to fool the underlying classification algorithm. So called \emph{adversarial images} embed noise into the image scenery to cause misclassification \cite{Machado2020AdversarialPerspective}. Given these potential vulnerabilities, AVs often incorporate some additional 'reflective' sensing such as LIDAR or ULTRASONICS to measure distance in real-time. However, these are also susceptible to attack.

\subsection{Active Sensors Attacks}
Resolving images in real-time for AVs requires substantial processing power and this is not always possible. Therefore, to ensure AVs can operate on orders of magnitude faster than real-time image processing, AVs will often incorporate Time-of-Flight (ToF) sensors such as LIDAR (travel) or ultrasonics for range detection. However, these methods of sensing can often be easily spoofed. For example, a ToF sensor for assisted braking device could easily be spuriously triggered by simply covering the sensor mid-journey i.e. with a simple bug controlled remotely. If timed appropriately this could cause traffic accidents or other outcomes with a relatively simple \emph{Sensor spoof} \cite{Cao2019AdversarialDriving}.

One of the best ways to prevent upstream sensing attacks is the corroboration of a sensor measurement by another measurement i.e. \emph{Sensor corroboration}. The deterministic relationship between different measurements can be used to examine or evaluate when sensors have been compromised or have embedded errors. Residual error detection used in  state estimation in power systems \cite{Monticelli2000ElectricEstimation} is an example of this approach. 

By using the expected relationships between local measurements minor errors can be overcome and erroneous measurements discarded. This approach is not full proof. Deception style attacks like \emph{False data injection} (FDI) attacks such as those outlined in \cite{Liu2011FalseGrids}. These attacks demonstrate that if an attacker is aware of the relationship between measurements, they can often structure their attack to bypass measurement corroboration techniques such as \emph{Residual error} or even \emph{Probabilistic detection}. However, this often requires the attacker to compromise multiple sensors simultaneously. A popular type of defence against deception-style attacks is \emph{Moving Target Defence} (MTD).

The fundamental issue with countermanding physical layer attacks is the lack of defensive capability when interpreting physical phenomena. \emph{Analytic-based} defences can help to address this. Analysis of inbound data measurements to ensure the inbound dataset is statistically reasonable given the constraints of reality. An example of this might be to examine inbound ToF sensor data and try to assess the potential acceleration rate based on the inbound data. 
\subsection{Side-Channel Attacks}
Side-channel attacks can be leveraged against the physical layer to exploit unintended information leakage from hardware components. As anyone can purchase an AV, the hardware components will have limited physical protections. Attackers maybe able to analyze physical phenomena to extract cryptographic keys or other confidential information by observing patterns that correlate with specific operations. 

\section{Cyber Layer Attacks}
The cyber-layer concerns what we might consider more conventional cyber-security concerns i.e. zero-day vulnerabilities, back-doors, and social engineering. Broadly, cyber layer vulnerabilities have been well considered with solutions such as encryption, password protection and information security. However, there are some key areas of cyber layer defence which we believe are less well considered and require new solutions.

\subsection{Code Vulnerabilities}
Coding languages can come with embedded vulnerabilities which an AV manufacturer should be cognizant of. C++ is a language commonly used in low-level programming for AVs and there are several attack types which emerge from the code base directly \cite{VerdiAnExamples}. These include (but are not limited to);  \emph{Code injections} (LDAP, Xpth, SQL etc.) whereby code is injected along with the normal query to access additional results. \emph{Memory corruption} attacks due to C++'s lack of spatial memory safety. \emph{Code re-use} attacks where code is repurposed to find arbitrary solutions. Attackers can also take advantage of data leakage from C++ in order to perform \emph{Denial of Service} (DoS) attacks.

There are several approaches to prevent bad code in AVs. Automotive coding standards such as the Motor Industry Software Reliability Association (MISRA) can formalise coding structures for code reviews \cite{DavidWard2006MISRASoftware}. By providing formal standards, it aims to minimise bugs and potential vulnerabilities. However, even when reviewed under formal standards code can still contain bugs which which may result in issues such as \emph{Buffer overflows} and \emph{Memory slippages}. RUST is comparatively safer
than C++ but can still be run in unsafe operating modes. Consequently, some projects have looked into how to formalize and ensure safe coding by design via a combination of in-memory architecture and coding structures. One such novel method is the Capability Hardware Enhanced RISC Instructions (CHERI) C/C++. CHERI C/C++ is a ground-up implementation aimed at providing fine-grained memory protection in coded C \cite{Richardson2020CompleteCapabilitiesb}. While low-level languages such as C are more susceptible, these kinds of faults, and memory vulnerabilities are not limited to C with JavaScript, Python and other popular languages having similar embedded security issues. 

\subsection{Adversarial Machine Learning}
Adversarial machine learning refers to a branch of software attacks specifically targeting the system's machine learning processes \cite{SivaKumar2020AdversarialPerspectives}. Broadly, these attacks can be classified into attacks against training data, attacks against the model itself and attacks against the federated system. \emph{Evasion attacks} are a form of model attack whereby the system is fed data in an attempt to create an error or misclassification in the machine learning model. A particularly relevant example of \emph{Evasion attacks} for AVs are adversarial images. 

\subsubsection{Adversarial Images}
Image processing is arguably the most important element of navigation for AVs. Image processing can be corrupted by alterations made to the underlying code base without the user being aware. Simple corruption of the image processing code base would likely render a vehicle inoperable. However, a worse outcome might occur when an attacker changes the underlying code base to misclassify events with the aim of creating dangerous scenarios. This could be done by changing the setup of the underlying classification algorithms or the image base upon which the classification is dependent. For example, swapping the classification of green and red light traffic light images could cause a car drive through a red light when it should be stopping. Defence against these attacks using code verification or version control should be trivial during times when the manufacturers' central authority can be reached. However, in the case of a self-driving vehicle there may be cases where the two code bases can't be compared (such as in a tunnel). This could represent an attacking opportunity for an adversary to intervene and produce differences between the local and established code bases. Therefore, strong tampering prevention techniques are needed when out of range of the central authority perhaps by locking down the code and making it read-only during non-communication. 

\subsubsection{Model Extraction}
Model extraction involves extracting the model training data which often has proprietary or commercial value. Often, with sufficient time and data gathering. It can be possible for an adversary to replicate a model outright from the data extracted. This can raise security concerns and increase the success rates for attacks like adversarial images. It can also have severe financial repercussions for the company if a sophisticated machine learning model can be replicated by a third party.

\subsubsection{Federated Systems Attacks}
Federated learning can be deployed in AVs allowing the algorithm to learn from multiple independent systems. For example, in the case of a fleet of self-driving cars you might seek to enhance the collision detection model by feeding back data from each individual car. The issue with this kind of learning model is the potential for Byzantine attacks. By capturing a large portion of the fleet, attackers can inject false data which maybe considered true if it is consistent with a large portion of the fleet. Federated systems can allow the adversary to perform data poisoning attacks by corrupting the underlying datasets.

\section{Communications Layer Attacks}
AVs place heavy reliance on fast and reliable network communications for getting real-time relevant information for continued operation (e.g. traffic data, street maps, GPS, weather information) or communicating with the cloud. There are $4$ types of communication networks involved, a) vehicle-to-vehicle (V2V), b) vehicle-to-infrastructure (V2I), c) vehicle-to-pedestrian (V2P) and d) vehicle-to-network (V2N). V2X (vehicle-to-everything) is a collective term used for the communication mechanism of autonomous vehicles. Each vehicle is installed with an on-board unit (OBU) for communication. 
Numerous network attacks are feasible for autonomous vehicles, and these security threats within the communication layer can be classified into distinct categories \cite{Meneguette2023VehicularChallenges}\cite{BrehonGrataloup2022MobileSurvey}.

\subsection{Confidentiality Attack}

Confidentiality is a key pillar of the CIA triad (Confidentiality, Integrity, Availability) in network security. Attackers can compromise confidentiality in various ways. Since the network is accessible to all vehicles, a malicious node can easily eavesdrop and capture sensitive information (\emph{Eavesdropping attack}). This can be done through \emph{Packet capturing}, where the attacker intercepts packets to extract confidential data. The attacker may then use this information to falsify location or traffic details. Weak encryption can also lead to \emph{Password attacks}, where attackers guess passwords using dictionaries or brute force. If an attacker gains access, it can compromise both the vehicle and traffic operations. Strong encryption and hashed passwords help mitigate these threats. Current solutions to manage \emph{Confidentiality attacks} often rely on public key infrastructure and secure communication protocols.

\subsection{Integrity Attack}
In this type of attack, the attacker compromises the integrity of the system by changing the transmitted data or traffic information. \emph{Replay attacks} are a commonly used form of this attack where the past message is replayed to gain the traffic network's trust. Also prominent are \emph{Masquerading attacks} and the \emph{Illusion attack}. In \emph{Masquerading attack}, an attacker masquerades a valid node's identity either by stealing or using some software bugs. By masquerading as a valid node, the attacker can intervene in the network causing various issues.

Conventional methods to address \emph{Integrity attacks} include integrity checks and digital signatures. However, it is possible that the key can be compromised and if an attacker gains access to the private key used for generating digital signatures, they can sign malicious data, making it appear genuine. Additionally, if an adversary gains physical or remote access to a vehicle's systems, they may tamper with data or manipulate digital signatures to compromise its integrity.

\subsection{Availability Attack}  

Attacking network availability is a common way to disrupt V2X communication. Traditional attacks like Denial of Service (DoS), Distributed Denial of Service (DDoS), spamming, and flooding remain significant threats. An attacker can flood the network with fake requests, exhausting resources and preventing legitimate traffic (DoS). Spamming and flooding increase delays, and when generated by botnets, they become harder to mitigate (DDoS).

Specific to autonomous vehicles (AVs), \emph{Blackhole} and \emph{Wormhole} attacks target availability. In a \emph{Blackhole attack}, a malicious node pretends to be the nearest node and drops packets, causing loss. In a \emph{Wormhole attack}, malicious nodes create a tunnel to intercept and modify information, creating false traffic scenarios.

AV networks are vulnerable to availability attacks that could disrupt critical systems like GPS, satellite communication, internet, and internal telemetry, leaving vehicles inoperable. Traditional countermeasures, such as redundancy, backup sensors, and failover systems, aim to ensure operation during disruptions. However, these may be inadequate against emerging threats exploiting vulnerabilities in interconnected AV networks.

\subsection{Identification Attack} Identification attacks strike on the node identity by creating false identities. The most common form of identification attack is a sybil attack where a single malicious node pretends as multiple nodes and transmits the wrong message in the network. A normal node follows the message as it seems to be received from multiple identities which may lead to traffic congestion or accidents. An attacker can also sniffs to get the key/certificate of a genuine node and later uses that to act as the node to transfer malicious message including malware and virus.

Attacks such as these can lead to delays in the transmission of important messages, which could lead to collision or other navigation issues. There is also the possibility that due to the authentication issues timing attacks can cause a genuine AV to be detected as a malicious entity not allowed to interact with the network. Addressing identification attacks involves implementing robust user authentication methods, such as biometrics and device fingerprinting, alongside continuous monitoring and behavior analysis to safeguard against impersonation and unauthorized access.

\subsection{Port Overwriting}
Hardware entry points such as serial ports provide very little native protection against direct intervention. Attacks on these entry points could allow an attacker to replicate many of the discussed attacks such as FDI or replay attacks. \emph{Man-in-the-middle} (MitM) are also possible via serial port interception whereby the attacker sits in between two systems and transmits the data between them. The data transmitted into and out of serial ports can lead to \emph{Side-channel} attacks. Whereby data is collected illegally via data leakage in industrial equipment.

If an attacker can replace the feed from a sensor or set of sensors they could reflect almost any system state they wanted to the user. \emph{Replay} or \emph{Triggering} attacks could be easily implemented without the level of intervention required in the physical sensing attacks. Further, because the access points for the serial ports for an AV are generally under the hood they might be harder to spot via a visual inspection of the AV.

\section{Cross-layer Attacks}

In addition to any vulnerabilities located in the respective layers, AVs can elucidate attack opportunities which cannot be captured in a single layer. Cross-layer attacks target multiple layers of the system simultaneously, utilise partial vulnerabilities in each domain and exploit the interactions between them which leads to complex and severe consequences. A few potential cross-layer attacks affecting the security of AVs are as follows.

\subsection{Sensor Manipulation and Adversarial Machine Learning} Cyber-Physical layer attack where an attacker manipulates the AV's physical environment to deceive its sensors while simultaneously exploiting vulnerabilities in the machine learning models used to process sensor data. For example, an attacker might utilise a scenery attack in combination with adversarial machine learning techniques, this could cause the vehicle to ignore critical road signs, leading to dangerous driving.

\subsection{False Data Injection and Network Tampering}
Physical-communications layer attack which involves injecting false data into the vehicle’s sensor system while simultaneously tampering with communication protocols. For example, an attacker could introduce false GPS signals (a physical layer attack) to mislead the vehicle about its actual location. Simultaneously, they could intercept and alter V2V communications (a communication layer attack) to provide corroborative but incorrect location data. This dual manipulation could cause the vehicle to take dangerous actions, such as making sudden turns or stops, which might lead to accidents or traffic disruptions.

\subsection{Side-Channel Attack with Code Injection} 
Cyber-Communications style attackers might use side-channel analysis to gather information about the vehicle’s internal operations. Once the information is extracted, it could be used to inject malicious code into the vehicle’s software through a communication interface. This malicious code could then be used to disable critical vehicle functions, disrupt communications, or even take control of the vehicle remotely. The combination of side-channel attacks with code injection through communication channels creates a potent cross-layer cyber-communications threat that severely compromises the vehicle’s safety and security.

\section{Conclusions \& Future Directions}
The AV cyber-security sector is rapidly evolving, with increasing emphasis on developing robust defense mechanisms to protect against sophisticated cyber threats. Industry leaders are prioritizing advanced encryption, secure communication protocols, and real-time threat detection to safeguard vehicle systems from potential breaches. Despite these efforts, the sector faces ongoing challenges in addressing vulnerabilities and ensuring comprehensive security as the technology continues to advance and integrate with other systems. While we have discussed a number of vulnerabilities and potential solutions here, we target the below subsections as priorities for increasing security within the cyber-physical and communications stack. 

\subsubsection{Spoof Proof Sensing} Sensors are some of the best targets for attacking an AV. Physical sensors possess few opportunities for upstream authentication and security. Current sensing modules are poorly defended with many cyber attack outcomes possible via the manipulation of the underlying measurement system. From simple brake triggering attacks to more complicated attacks such as adversarial images or false data injection. It is clear that we need a framework for guaranteeing high-quality sensing protection.

\subsubsection{Secure Entry Points}
Weak entry points provide an attacker with multiple options for attack. From MitM attacks, side-channel attacks, FDI and replay attacks the standard protocols employed by many of the entry points on an AV offer easy attack opportunities to attackers. The distributed nature of entry points, especially across large networks makes developing generic solutions difficult. In the future, the vulnerability of low-security access points needs to be considered and appropriate defences incorporated into the system stack. These defences can be split broadly into pre and post-intrusion defence solutions. Pre-intrusion considers those forms of protection which prevent the attacker from gaining access through the port. 
Post-intrusion defences refer to catching attackers based on their activities once they have access to the system such as through analytics or moving target style detection schemes.


\subsubsection{Software Secure by Design}
A more conventional target base, the software in an AV needs to be secure by design from attack. At the coding level, secure-by-design coding tools such as the CHERI architecture or languages like RUST can help prevent common code vulnerabilities in memory. Alterations to the classification base can also be devastating for AVs making them unable or dangerous to operate. Ideally, we would develop local and centralized solutions to this problem to ensure compliance with a good classification base even when the central authority is unavailable. 

\subsubsection{Operational Considerations}
Enhancements to AV security in relation to their operational environment are also critical. Operators should consider that AV security extends beyond the domain of the vehicle into the surrounding operational environment. Considerations on how to mitigate these risks should be made. 

\subsubsection{Post-Quantum Future Proofing}
AVs are an emergent technology and the security challenges that face them will also need to be future-proofed. As viable quantum computers come online, system operators will need to address the vulnerability of conventional encryption approaches. To mitigate the security threats possessed by quantum computers, NIST has approved a few post-quantum secured algorithms \cite{Moody2020StatusProcess}. However, the wide adoption of these algorithms has many challenges. Different post-quantum algorithms might allow trade-offs between the processing time and key/signature size, which might influence the selection of the algorithm to use. Some of the algorithms are also vulnerable to side-channel attacks which need to be managed.



\section*{Acknowledgment}
This work was supported by UK Research and Innovation’s Digital Security by Design (DSbD) Grant and was performed in conjunction with CyberHive.

\ifCLASSOPTIONcaptionsoff
  \newpage
\fi






\section*{Biography}
\textbf{Martin Higgins} (S’19) is a research associate at the University of Oxford on the Digital Security by Design Project.  His research interests lie in power systems, cyber-security, false data injection attacks, and autonomous vehicles. Martin received his PhD in electrical engineering from Imperial College in 2021. Contact him at Martin.Higgins@eng.ox.ac.uk.

\textbf{Devki Nandan Jha} is a  Lecturer at Newcastle University, UK. He is also a visiting researcher at the Oxford e-Research Centre, University of Oxford, UK. He has a PhD in Computer Science from Newcastle University, Newcastle Upon Tyne, UK. His research interests include cloud computing, internet of things, trust and security, and machine learning. Contact him at Dev.Jha@ncl.ac.uk. 

\textbf{David Blundell} is the founder, MD and CTO of CyberHive and 100 Percent IT Ltd. His current research interests are secure distributed computing, autonomous vehicles and post-quantum encryption. Contact him at David.Blundell@cyberhive.com.

\textbf{David Wallom} is an Associate Director of Innovation with the Oxford e-Research Centre, University of Oxford. His current research interests include applications and reuse of e-infrastructure, as well as the application of high-performance computing techniques and cyber-security. Contact him at David.Wallom@oerc.ox.ac.uk.


\end{document}